\documentclass[11 pt, a4paper]{article}
\usepackage{jheppub}
\usepackage{amsmath}

\newcommand{\Tr}{{\rm Tr\;}}

\usepackage{graphicx}
\usepackage{amsmath}
\usepackage{amsfonts}
\usepackage{epsfig}
\usepackage{slashed}
\usepackage{braket}

\usepackage{comment}
\usepackage{cases}

\def\nn{\nonumber}
\def\bec{\begin{center}}
\def\eec{\end{center}}
\def\beq{\begin{equation}}
\def\eeq{\end{equation}}
\def\bea{\begin{eqnarray}}
\def\eea{\end{eqnarray}}

\def\ahat{\hat{a}}
\def\bhat{\hat{b}}
\def\chat{\hat{c}}
\def\dhat{\hat{d}}
\def\ehat{\hat{e}}

\newcommand{\cF}{{\cal F}}
\newcommand{\cFb}{{\overline{\cal F}}}
\newcommand{\cD}{{\cal D}}
\newcommand{\cDb}{{\overline{\cal D}}}
\newcommand{\cQ}{{\cal Q}}
\newcommand{\cU}{{\cal U}}
\newcommand{\cN}{{\cal N}}
\newcommand{\cUb}{{\overline{\cal U}}} 

\title{Supersymmetric Lattice Theories: Contribution to Snowmass 2022}
\author[a]{Simon Catterall and}
\author[b]{Joel Giedt}

\affiliation[a]{Department of Physics, Syracuse University, Syracuse, NY 13244, USA }
\affiliation[b]{Department of Physics and Astronomy, RPI, Troy, NY 12180, USA}

\abstract{In this white paper we summarise the construction and applications of 
lattice theories possessing exact supersymmetry focusing, in particular, on $\cN=4$ Yang-Mills theory.
Lattice formulations of this theory allow for numerical simulation of the theory
at strong coupling and hence give a window on non-perturbative physics away from the planar limit. This has important applications to our understanding of holographic
approaches to quantum gravity and conformal field theories.
}

\begin{document}
\maketitle

\section{Executive summary}
In this white paper we summarise the construction and applications of 
lattice theories possessing exact supersymmetry focusing, in particular, on $\cN=4$ Yang-Mills theory.
Lattice formulations of this theory allow for numerical simulation of the theory
at strong coupling and hence give a window on non-perturbative physics away from the planar limit. This has important applications to our understanding of holographic
approaches to quantum gravity and conformal field theories.
In particular:
\begin{itemize}
    \item We find that quantities scale with the 't Hooft coupling $\lambda$ in a way that is consistent with holography.  In particular, Wilson loops scale as $\exp(- c \sqrt{\lambda})$, where $c$ is some constant.
    \item Success in this regime opens the door to other interesting studies at strong coupling and away from the planar limit
    including tests of S-duality, computations of
    the dimension of the Konishi operator and calculations of string loop
    corrections to classical supergravity.
    \item Such calculations can also help bridge to other theoretical efforts such as
    the scattering amplitudes and
    conformal bootstrap programs.
\end{itemize}

\section{Review}
In recent years a new approach to the problem of formulating supersymmetric lattice
theories has been developed with the result that a certain class of supersymmetric theory can be discretized while preserving one or more supercharges at non-zero lattice
spacing. These theories can be derived in two independent ways; by exploiting orbifold
and deconstruction techniques or by careful discretization of a topologically twisted
formulation of the target supersymetric theory \cite{Catterall:2009it}~\footnote{Actually the orbifold methods
only yield Yang-Mills theories while the topological constructions are also capable
of describing Wess-Zumino models}.

In the case of $\cN=4$ Yang-Mills the resultant lattice action is 
\beq
S=\frac{N}{4\lambda} \cQ \sum_{x}\Tr \left(\chi_{ab}\cF_{ab}+\eta \cDb_a\cU_a+\frac{1}{2}\eta d+
\kappa\,\eta\,\left({\rm Re\,det}\left[\cU_a(x)\right]-1\right)\right)+S_{\rm closed}
\eeq
where the lattice field strength
\beq\cF_{ab}(x)=\cU_a(x)\cU_b(x+\ahat)-\cU_b(x)\cU_a(x+\bhat)\eeq where $\cU_a(x)$
denotes a {\it complexified} gauge field living on the lattice link running from $x\to x+\ahat$ and where $\ahat$ denotes one
of the five basis vectors of an underlying $A_4^*$ lattice.
Similarly
\beq\cDb_a \cU_a=\cU_a(x)\cUb_a(x)-\cUb_a(x-\ahat)\cU_a(x-\ahat).\eeq
The five fermion fields $\psi_a$, being superpartners of the gauge fields, live on the 
corresponding links, while
the ten fermion fields $\chi_{ab}(x)$ are associated with new face links running
from $x+\ahat+\bhat\to x$. The scalar fermion $\eta(x)$ lives on the lattice site $x$ and is associated with a conserved supercharge $\cQ$
which acts on the fields in the following way\footnote{One of the things that is learned from the orbifold construction is that the number conserved supercharges is equal to the the number of site fermions.}
\begin{align}
\cQ\, \cU_a&\to \psi_a\nn\\
\cQ\, \psi_a&\to0\nn\\
\cQ\, \eta&\to d\nn\\
\cQ\, d&\to 0\nn\\
\cQ\, \chi_{ab}&\to \cFb_{ab}\nn\\
\cQ\, \cUb_a&\to 0
\end{align}
Notice that $\cQ^2=0$ which guarantees the supersymmetric invariance of the
first part of the lattice action. The auxiliary site field $d(x)$ is needed for nilpotency of $\cQ$ offshell. The second term $S_{\rm closed}$ is given by
\beq
S_{\rm closed}=-\frac{N}{16\lambda}\sum_x \Tr \epsilon_{abcde}\chi_{ab}\cDb_c\chi_{de}\eeq
where the covariant difference operator acting on the fermion field $\chi_{de}$ takes the form
\beq
\cDb_c\chi_{de}(x)=\cUb_c(x-\chat)\chi_{de}(x+\ahat+\bhat)-\chi_{de}(x-\dhat-\ehat)\cUb_c(x+\ahat+\bhat)\eeq
To retain exact supersymmetry all fields reside in
the algebra of the gauge group -- taking their values in the adjoint representation of $U(N)$:
$f(x)=\sum_{A=1}^{N^2} T^A f^A(x)$ with $\Tr (T^A T^B)=-\delta^{AB}$.
The latter term can be shown to be supersymmetric via an exact lattice
Bianchi identity $\epsilon_{abcde}\cDb_c \chi_{de}=0$. This action is invariant
under $\cQ$, $SU(N)$ lattice gauge invariance and the $S^5$ point group symmetry of the
$A_4^*$ lattice.\footnote{Notice that there are five lattice vectors, ${\hat a} = {\hat 1}, \ldots, {\hat 5}$, corresponding to the nearest-neighbor links of the $A_4^*$ lattice, and the fact that we have five complexified ``gauge fields.''  The $A_4^*$ lattice is four-dimensional, in spite of having five primitive vectors.}
Carrying out the $\cQ$ variation and integrating out the auxiliary field $d$ we obtain the
supersymmetric lattice action $S=S_b+S_f$ where
\beq
S_b=\frac{N}{4 \lambda} \sum_x \Tr \left( \cF_{ab} \cFb_{ab} \right)
+ \frac{1}{2} \Tr \left( \cDb_a \cU_a-\kappa \left[ {\rm Re \, det} \left[\cU_a(x)\right]-1\right]^2\right)\eeq
and
\beq
S_f=-\frac{N}{4\lambda}\sum_x \left(\Tr\chi_{ab}\cD_{\left[a\right.}\psi_{\left. b\right]}+
\eta \cDb_a\psi_a-\frac{\kappa}{2}\Tr(\eta){\rm det\,} (\cU_a(x))\Tr(\cU_a^{-1}(x)\psi_a(x))\right)+S_{\rm closed}\eeq
In the continuum this action can be obtained by discretization of
the Marcus or GL twist of $\cN=4$ Yang-Mills 
but in flat space is completely equivalent to it. In the continuum the twist is done as a prelude
to the construction of a topological quantum field theory but in the context of
lattice supersymmetry it is merely used as a change of variables that
allows for discretization while preserving a single exact supersymmetry. The twisting removes the spinors
from the theory replacing them by the antisymmetric tensor fields $\eta,\psi_a,\chi_{ab}$
which appears as components
of a K\"{a}hler-Dirac field. The latter is equivalent at zero coupling to a (reduced) staggered field
and hence describes four physical Majorana fermions in the continuum limit - as required for $\cN=4$ Yang-Mills. The
twisting procedure also
packs the six scalar fields of the continuum theory together with the four gauge fields
into five complex gauge fields corresponding to the lattice fields $\cU_a$. The coupling $\kappa$ is needed to project the theory from $U(N)$ to $SU(N)$ and thereby evade instability
issues that otherwise would arise at strong coupling.

General arguments have been put forward that the theory should approach the continuum
${\cal N}=4$ theory after tuning a single marginal operator \cite{Catterall:2013roa}. The theory can
be simulated using the same algorithms that are employed for lattice QCD
\cite{Catterall:2011pd,Catterall:2012yq,Catterall:2014vka}.~\footnote{The theory does not appear to suffer from a sign problem although the exact reasons for
this are not well understood \cite{Catterall:2020lsi}.}
It has also been used to explore the physics of black holes
and gauge-gravity duality in lower dimensions~
\cite{Anagnostopoulos:2007fw,Hanada:2008gy,Catterall:2008yz,Catterall:2009xn,Catterall:2010fx,Hanada:2016zxj,Berkowitz:2016jlq,Catterall:2017lub,Rinaldi:2017mjl, Catterall:2020nmn}.
There is one final wrinkle that needs to be mentioned. To regulate the flat directions
of the theory to do simulations it is necessary to add
a soft supersymmetry breaking  term of the form
\beq
S_{\rm mass}=\mu^2\sum_x \Tr\left(\cUb_a(x)\cU_a(x)-I\right)^2\eeq
While this breaks the exact supersymmetry softly all counter terms induced by this
breaking will have couplings that are multiplicative in $\mu^2$  and hence vanishing
as $\mu^2\to 0$. 

\section{Conformal invariance and holography}

$\cN=4$ Yang-Mills is thought to be a non-trivial conformal field theory
for any value of the 't Hooft coupling. Simulations are consistent with this and
show a single phase theory with vanishing string tension. Furthermore, the theory
can be solved in the planar limit $N\to\infty$ and
exhibits a non-trivial dependence on the 't Hooft coupling $\lambda$. Specifically circular
supersymmetric Wilson loops $W_{\rm susy}$ in the planar strong coupling limit
are independent of size and depend only
on $\sqrt{\lambda}$ \cite{Drukker:2000rr,Maldacena:1998im}.
\[ \ln {W_{\rm susy}} = {\rm const} \sqrt{\lambda} \]
This result was first derived by exploiting holography to relate this Yang-Mills theory
to classical supergravity in five dimensional $AdS$  space.

\begin{figure}[htbp]
\centering
\includegraphics[width=0.75\textwidth]{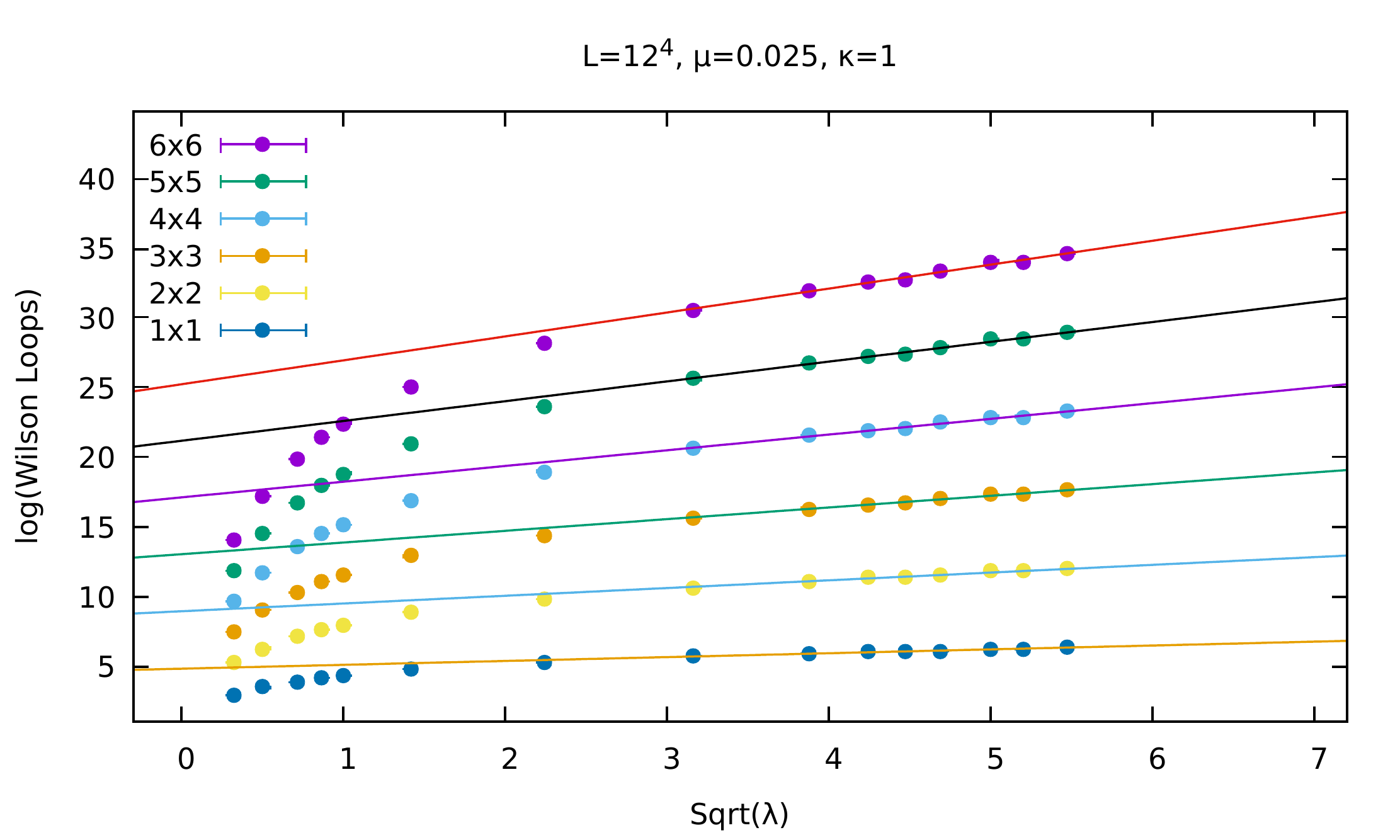} 
\caption{\label{loops}Supersymmetric $n\times n$ Wilson loops on $12^4$ lattice at $\mu=0.025$ }
\end{figure}

\begin{figure}[htbp]
\centering
\includegraphics[width=0.75\textwidth]{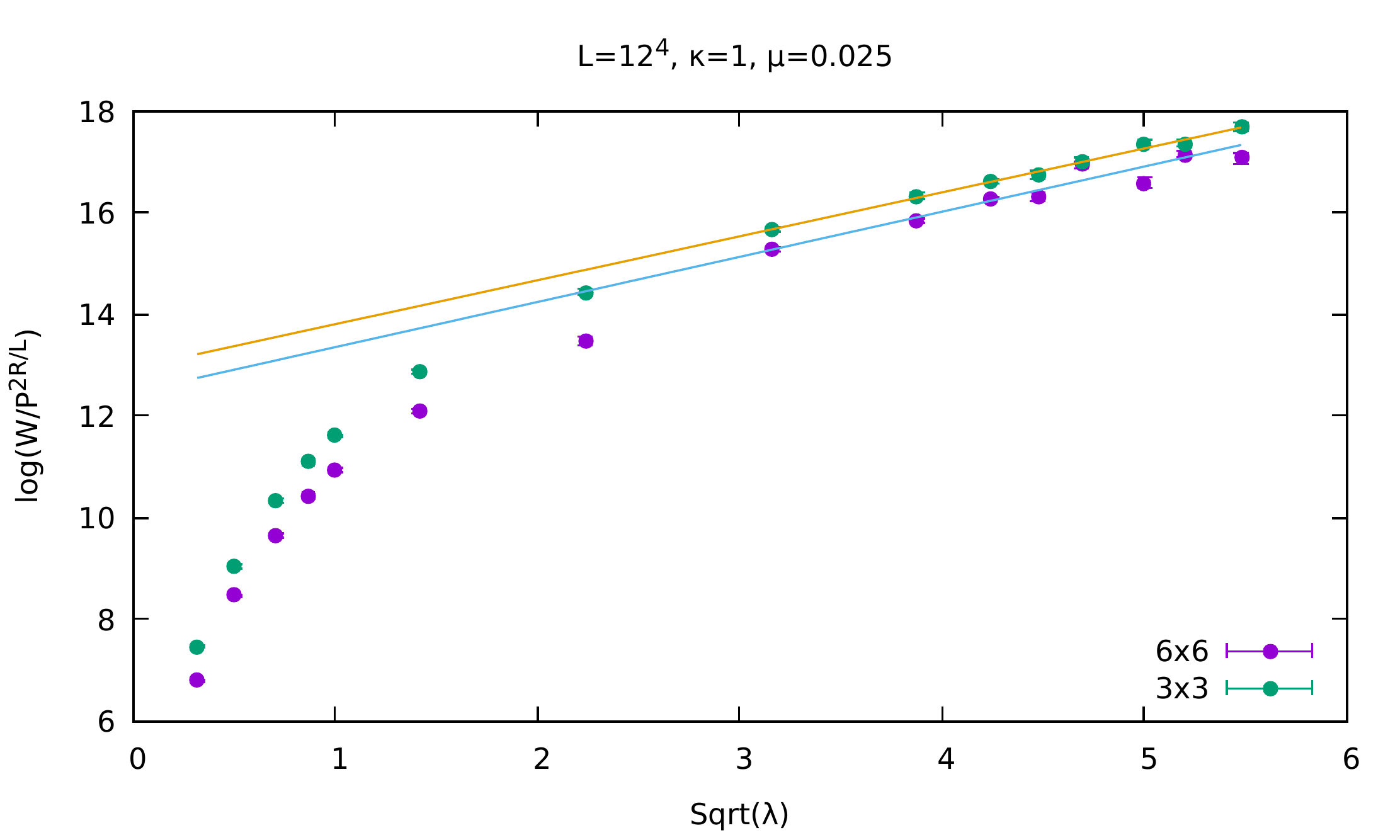} 
\caption{\label{wilson6}Renormalized supersymmetric $6\times 6$ and $3\times 3$ Wilson loops on $12^4$ lattice at $\mu=0.025$ }
\end{figure}

The characteristic $\sqrt{\lambda}$ dependence can also be seen in the results
of numerical simulations at strong coupling {\it even for small numbers of
colors} - see fig.~\ref{loops} which plots the logarithm of the square lattice Wilson loop constructed from $\cU_a$ as a function of $\sqrt{\lambda}$ for $N=2$.
The dependence on loop size $R$ reflects the presence of a constant perimeter term in
the static potential arising from the (static) quark mass \cite{Erickson:2000af}. Indeed if this is subtracted out by
normalizing the Wilson loops by appropriate powers of the Polyakov line one obtains
the plot in fig.~\ref{wilson6} which exhibits both an insensitivity to
loop size and also the $\sqrt{\lambda}$ 
behavior expected from holography. 
The strange $\sqrt{\lambda}$ dependence {\it cannot} be seen in perturbation theory
and this result is a very non-trivial test of the correctness of the lattice approach in a non-perturbative regime.

\section{Future Directions - executive summary}
Supersymmetric lattice actions can be formulated which conserve one or more continuum supersymmetries and flow to the continuum theory with minimal tuning as the lattice
spacing is sent to zero. One of the most interesting examples that has been studied
is $\cN=4$ super Yang-Mills. Results that have been obtained so far are consistent with
a single conformally invariant phase for any value of the 't Hooft coupling
and agree with holographic predictions for Wilson
loops even for small numbers of colors -- an unexpected
and non-trivial result. Future work will focus on
a variety of outstanding issues
\begin{itemize}
    \item Look for precise numerical agreement of the lattice and
    continuum results for supersymmetric Wilson loops in the planar limit at strong coupling.
    \item Explore whether fine tuning is indeed needed to restore the remaining supersymmetries
    in the continuum limit.
    \item Compute the Konishi operator and supergravity operator scaling dimensions that characterize
    the conformal behavior of the theory for arbitary numbers of colors comparing with
    bootstrap and planar calculations.  Here it is important to take into account the impact of discretization on the $SU(4)_R \simeq SO(6)$ flavor symmetry.
    \item Search for evidence of S-duality in the lattice theory by measuring
    gauge boson and monopole masses in the Coulomb phase of the lattice theory.  Here one has a precise, BPS-protected formula to compare to.
\end{itemize}

\acknowledgments
This work was supported by the US Department of Energy (DOE), Office of Science, Office of High Energy Physics, 
under Award Numbers {DE-SC0009998} (SC) and {DE-SC0013496} (JG). Numerical calculations were carried out on the DOE-funded USQCD facilities at Fermilab.

\bibliographystyle{JHEP3}
\bibliography{susy}

\end{document}